\documentclass[iop]{emulateapj}
\usepackage{amsmath}

\newcommand{\bmath}[1]{\mbox{\boldmath{$#1$}}}

\newcommand{\del}{{\bf \nabla}}

\newcommand{\cs}{c_{\rm s}}

\begin{document}

\title{Efficiency of Particle Trapping in the Outer Regions of Protoplanetary Disks}

\author{Jacob B. Simon\altaffilmark{1,2,3} and Philip J. Armitage\altaffilmark{3,4}}

\email{jbsimon.astro@gmail.com}

\begin{abstract}
We investigate the strength of axisymmetric local pressure maxima (zonal flows) in the outer regions of
protoplanetary disks, where ambipolar diffusion reduces turbulent stresses driven by the magnetorotational
instability. Using local numerical simulations we show that in the absence of net vertical magnetic fields, the
strength of turbulence in the ambipolar dominated region of the disk is low and any zonal flows that are present are weak.
For net fields strong enough to yield observed protostellar accretion rates, however, zonal flows with a 
density amplitude of 10-20\% are formed. These strengths are comparable to those seen in simulations of ideal MHD disk
turbulence. We investigate whether these zonal flows are able to reverse the inward radial drift of solids, leading to
prolonged and enhanced concentration as a prelude to planetesimal formation. For commonly assumed mean surface density profiles (surface density
$\Sigma \propto r^{-1/2}$ or steeper) we find that the predicted perturbations to the background disk profile 
do not correspond to local pressure
maxima. This is a consequence of radial width of the simulated zonal flows, which is larger than was 
assumed in prior analytic models of particle trapping. These larger scale flows would only trap particles 
for higher amplitude fluctuations than observed.
We conclude that zonal flows are likely to be present in the outer regions of
protoplanetary disks and are potentially large enough to be observable, but are unlikely 
to lead to strong particle trapping.
\end{abstract} 


\altaffiltext{1}{Department of Space Studies, Southwest Research Institute, Boulder, CO 80302}
\altaffiltext{2}{Sagan Fellow}
\altaffiltext{3}{JILA, University of Colorado and NIST, 440 UCB, Boulder, CO 80309-0440}
\altaffiltext{4}{Department of Astrophysical and Planetary Sciences, University of Colorado, Boulder, CO 80309}

\section{Introduction} 
The radial drift problem in protoplanetary disks results from aerodynamic drag between particles and 
gas, which orbit at slightly different speeds due to pressure gradients in the gas \citep{weidenschilling77}. 
Drift is fastest for particles whose surface area to mass ratio is such that they are marginally coupled, 
with a dimensionless stopping time $\tau = m \Delta v \Omega^{-1} / |F_{\rm drag}| \sim 1$. In the outer 
regions of protoplanetary disks, at 30 to 100~AU from the star, this condition is met for spherical particles 
with size $\sim 1 \ {\rm cm}$. Particles of roughly similar size -- including mm-sized particles that can be directly 
observed in dust continuum observations -- would therefore be expected to drain rapidly from the outer 
disk, leaving gas behind \citep{youdin04,hughes12}. In fact, although the extremities of 
some well-studied disks {\em are} gas-rich \citep{andrews12}, mm-sized particles appear 
to be retained in the outer disk \citep{ricci10} for substantially longer than is theoretically predicted 
\citep{brauer07}. A candidate explanation is that dust is trapped in either permanent or transient 
local pressure maxima that act to slow radial drift driven by the global pressure gradient \citep{pinilla12,dittrich13}. 
If such traps are present, the resulting over-density of solids could locally promote the growth of 
instabilities leading to planetesimal formation \citep{chiang10}.  

In this paper, we show that under the physical conditions appropriate to the outer regions of protoplanetary disks,
zonal flows form spontaneously within magnetohydrodynamic (MHD) disk turbulence. Zonal
flows are  transient axisymmetric pressure maxima that were seen in local ideal MHD simulations of accretion  disks
\citep{johansen09}. They are sustained by a geostrophic balance between  pressure gradients and
Coriolis forces. Subsequent work has shown zonal flows to be a robust outcome of  ideal MHD disk turbulence in the
local limit \citep{simon12,dittrich13} and has identified similar  structures in global simulations
\cite[e.g.,][]{dzyurkevich10,flock11,uribe11}. What has not been established is  whether, in the presence of the
non-ideal MHD effects (Ohmic and ambipolar diffusion, and the  Hall effect) that are important in protoplanetary
disks \citep{armitage11}, zonal flows remain  strong enough to act as efficient traps. Here, we address this question
for the outer disk,  where ambipolar diffusion creates an ``ambipolar damping zone"
\citep{perezbecker11,mohanty13,simon13a,simon13b}.  

Our results are based on local (shearing-box) simulations of non-ideal MHD disk turbulence, using methods described 
in Section 2. In Section 3 we show that within the ambipolar damping zone, the amplitude of predicted zonal flows 
is a function of the net flux of magnetic field that threads the disk. Zero net flux simulations yield weak 
turbulence \citep{simon13a} and very weak zonal flows, while simulations with net fields chosen to yield the observed accretion
rates onto young stars \citep{simon13b} produce prominent zonal flows. In Section 4 we explore  
whether the derived zonal flows would lead to particle trapping, under the assumption that the perturbed global 
disk profile can be approximated as the mean global profile multiplied by the local perturbed structure. 
In Section 5 we compare our results
with previous work and discuss the general implications of our results. 

\section{Methods}
We simulate the evolution of the magnetorotational instability \cite[MRI;][]{balbus98} within a  local, shearing-box domain \citep{hawley95},
including the effects of vertical stratification and ambipolar  diffusion. Overall, our calculations are similar to those reported in 
\citep{simon13b}. The equation of state is isothermal, and we employ vertical outflow
boundary conditions, modified to enhance the buoyant removal of magnetic flux from the domain. The shearing boxes include the physical 
effects expected to dominate at 
large radial distances from the central star and use a highly simplified ionization model in which a thin layer above and below the disk
mid-plane is assumed to be very strongly ionized due to stellar FUV photons \citep{perezbecker11}; below these highly ionized layers, we
assume a constant, yet large value for the strength of ambipolar diffusion. We describe the numerical details of the simulations below.

\subsection{Numerical Method}
\label{num_method}

We use \textit{Athena}, a second-order accurate Godunov
flux-conservative code for solving the equations of MHD. 
\textit{Athena} uses the dimensionally unsplit corner transport upwind (CTU) method
of \cite{colella90} coupled with the third-order in space piecewise
parabolic method (PPM) of \cite{colella84} and a constrained transport
\citep[CT;][]{evans88} algorithm for preserving the $\del \cdot {\bmath
B}$~=~0 constraint.  We use the HLLD Riemann solver to calculate the
numerical fluxes \cite[]{miyoshi05,mignone07b}.  A detailed description
of the base \textit{Athena} algorithm and the results of various test problems
are given in \cite{gardiner05a}, \cite{gardiner08}, and \cite{stone08}.

We take advantage of the shearing box approximation in order to better resolve
small scales where ambipolar diffusion becomes important.  The shearing box is a model
for a local, co-rotating disk patch whose size is small compared to the
radial distance from the central object, $R_0$.  This allows the
construction of a local Cartesian frame $(x,y,z)$ that is defined in terms of the disk's
cylindrical co-ordinates $(R,\phi,z^\prime)$ via  $x=(R-R_0)$, $y=R_0 \phi$, and $z = z^\prime$.
The local patch  co-rotates with an angular velocity $\Omega$ corresponding to
the orbital frequency at $R_0$, the center of the box; see \cite{hawley95}.  The equations to solve are:

\begin{equation}
\label{continuity_eqn}
\frac{\partial \rho}{\partial t} + \del \cdot (\rho {\bmath v}) = 0,
\end{equation}
\begin{equation}
\label{momentum_eqn}
\begin{split}
\frac{\partial \rho {\bm v}}{\partial t} + \del \cdot \left(\rho {\bm v}{\bm v} - {\bm B}{\bm B}\right) + \del \left(P + \frac{1}{2} B^2\right) \\
= 2 q \rho \Omega^2 {\bm x} - \rho \Omega^2 {\bm z} - 2 {\bm \Omega} \times \rho {\bm v} \\
\end{split}
\end{equation}
\begin{equation}
\label{induction_eqn}
\frac{\partial {\bmath B}}{\partial t} - \del \times \left({\bmath v} \times {\bmath B}\right) = \del \times \left[\frac{({\bmath J}\times {\bmath B})\times {\bmath B}}{\gamma \rho_i \rho}\right],
\end{equation} 

\noindent 
where $\rho$ is the mass density, $\rho {\bmath v}$ is the momentum
density, ${\bmath B}$ is the magnetic field, $P$ is the gas pressure,
and $q$ is the shear parameter, defined as $q = -d$ln$\Omega/d$ln$R$.
We use $q = 3/2$, appropriate for a Keplerian disk.  For simplicity and numerical convenience, we
assume an isothermal equation of state $P = \rho \cs^2$, where $\cs$
is the isothermal sound speed.  From left to right, the source terms
in equation~(\ref{momentum_eqn}) correspond to radial tidal forces
(gravity and centrifugal), vertical gravity, and the Coriolis force. The source
term in equation~(\ref{induction_eqn}) is the effect of ambipolar diffusion
on the magnetic field evolution, where $\rho_i$ is the ion density, and
$\gamma$ is the coefficient of momentum transfer for ion-neutral
collisions. Note that our system of units has the magnetic permeability $\mu = 1$, and
the current density is

\begin{equation}
\label{current}
{\bmath J} = \del \times {\bmath B}.
\end{equation}

\noindent Numerical algorithms for integrating these equations are described in detail in
\cite{stone10} (see also the Appendix of \citealp{simon11a}). 
The $y$ boundary conditions are strictly periodic, whereas the $x$ boundaries
are shearing periodic \cite[]{hawley95}. The vertical boundary conditions are the
modified outflow boundaries described in \cite{simon13b}. The electromotive forces (EMFs) at
the radial boundaries are properly remapped to guarantee that the net
vertical magnetic flux is strictly conserved to machine precision using CT
\citep{stone10}. 

As in \cite{simon13a} and \cite{simon13b}, ambipolar diffusion is implemented in a first-order operator-split
manner using CT to preserve the divergence free condition
with an additional step of remapping $J_y$ at radial shearing-box boundaries.
The super time-stepping (STS) technique of \cite{alexiades96} has been implemented to
accelerate our calculations (see the Appendix of \cite{simon13a}).

\subsection{Am Profiles}
\label{am_profiles}

The strength of ambipolar diffusion is characterized by the
ambipolar Elsasser number
\begin{equation}
\label{am1}
{\rm Am}\equiv\frac{\gamma\rho_i}{\Omega},
\end{equation}

\noindent
which corresponds to the number of times a neutral molecule collides with the ions in a
dynamical time ($\Omega^{-1}$). The structure of the ambipolar damping zone is 
determined by the vertical profile of Am, which depends upon the assumed disk 
model, and on the strength of magnetic fields that are self-consistently present 
within the disk.

The ionization structure that we employ is motivated by the same arguments and assumptions made in \cite{simon13b}. We adopt the minimum-mass
solar nebular (MMSN) disk model with $\Sigma=1700R_{\rm AU}^{-3/2}$g cm$^{-2}$ \citep{weidenschilling77,hayashi81}, where $R_{\rm AU}$ is the
disk radius measured in AU. We choose the Am profile based on the far ultraviolet (FUV) ionization model of \citet{perezbecker11}, in which FUV
photons strongly ionize a column density of $\Sigma_i \sim0.01-0.1$~g~cm$^{-2}$. The corresponding value of Am within the FUV ionized layer
can be expressed as \citep{bai13b}, 

\begin{equation}
\label{Am_FUV}
{\rm Am_{\rm FUV}} \approx3.3\times10^7
\bigg(\frac{f}{10^{-5}}\bigg)\bigg(\frac{\rho}{\rho_{0,{\rm mid}}}\bigg)R_{\rm AU}^{-5/4}\ ,
\end{equation}

\noindent
where $f$ is the ionization fraction and $\rho_{0, {\rm mid}}$ is the mid-plane density.
For simplicity, we fix $f=10^{-5}$ and use an ionization depth of $\Sigma_i = 0.1$ g cm$^{-2}$.
We identify the location of the base of the FUV ionization layer ($z_t$ and $z_b$ for top and
bottom, respectively) by integrating at each time step the horizontally averaged mass density
from the boundary towards the mid-plane until $\Sigma_i$ is reached. We then
use Equation (\ref{Am_FUV}) to set the strength of ambipolar diffusion in the
ionized surface layers of the disk. In the mid-plane region $(z_b<z<z_t)$, we simply set Am~=~1, which leads
to an ``ambipolar damping zone" \citep{simon13b}.

Adopting this model, the value of Am changes quite dramatically from Am~=~1
to Am $\sim 10^4$ at the base of the FUV layer. This very large transition
is smoothed over roughly 7 grid zones so as to prevent a discontinuous transition in Am.
The smoothing functions we apply are based upon the error function (ERF).   Thus, the
complete profile of Am for these runs is given by

\begin{equation}
\label{amc}
\small
{\rm Am} \equiv \left\{ \begin{array}{ll}
 {\rm Am_{\rm FUV}} & \quad 
\mbox{$z \ge z_t + \Delta z$} \\
1 + \frac{1}{2}{\rm Am_{\rm FUV}}S^+(z)  & \quad
\mbox{$z_t - n\Delta z < z < z_t + \Delta z$} \\
1 & \quad
 \mbox{$z_b+n\Delta z \le z \le z_t - n \Delta z$} \\
1 + \frac{1}{2}{\rm Am_{\rm FUV}}S^-(z) & \quad
\mbox{$z_b-\Delta z < z < z_b + n\Delta z$} \\
{\rm Am_{\rm FUV}} & \quad
\mbox{$z \le z_b - \Delta z$}
\end{array} \right.
\end{equation}

\noindent
where $S^+(z)$ and $S^-(z)$ are the smoothing functions defined as 
 
\begin{equation}
\small
\label{splus}
S^+(z) \equiv 1+{\rm ERF}\left(\frac{z-0.9z_t}{\Delta z}\right),
\end{equation}
\begin{equation}
\small
\label{sminus}
S^-(z) \equiv 1-{\rm ERF}\left(\frac{z-0.9z_b}{\Delta z}\right). 
\end{equation}

\noindent
Here, $n = 8$ and $\Delta z = 0.05 H$.  These numbers were chosen to give a reasonably
well-resolved transition region between Am = 1 and ${\rm Am_{\rm FUV}}$. We note that
since Am$_{\rm FUV}\gg1$ in the above formula, the FUV photons effectively penetrate slightly
deeper than $z_t$ and $z_b$ by about $0.2H$.

\begin{widetext}
\begin{deluxetable*}{l|ccccc}
\tabletypesize{\small}
\tablewidth{0pc}
\tablecaption{Shearing Box Simulations\label{tbl:sims}}
\tablehead{
\colhead{Label}&
\colhead{Ionization Structure}&
\colhead{Vertical Magnetic Flux}&
\colhead{$\alpha_{\rm midplane}$}&
\colhead{$\alpha$}&
\colhead{$A$\tablenotemark{$\ast$}} }
\startdata
NF30AU & Ionization at 30 AU & Net vertical field with $\beta_0 = 10^4$ & 0.0017 & 0.018 & 0.17 \\
ZNF30AU & Ionization at 30 AU  & Zero net vertical field & $7.6\times10^{-6}$ & $6.8\times10^{-4}$ & 0.0031 \\
ZNF100AU & Ionization at 100 AU & Zero net vertical field & $8.4\times10^{-5}$ & 0.0017 & 0.015 \\
IDEAL & Fully ionized & Net vertical field with $\beta_0 = 10^4$ & 0.036 & 0.072 & 0.14 \\
\enddata
\tablenotetext{$\ast$}{\scriptsize This quantity is as defined in the text: the maximum of $|\overline{\delta\rho}_{\rm frac}|$}
\end{deluxetable*}
\end{widetext}

\subsection{Simulations}
\label{simulations}

\begin{figure*}[ht]
\begin{center}
\includegraphics[width=0.94\textwidth,angle=0]{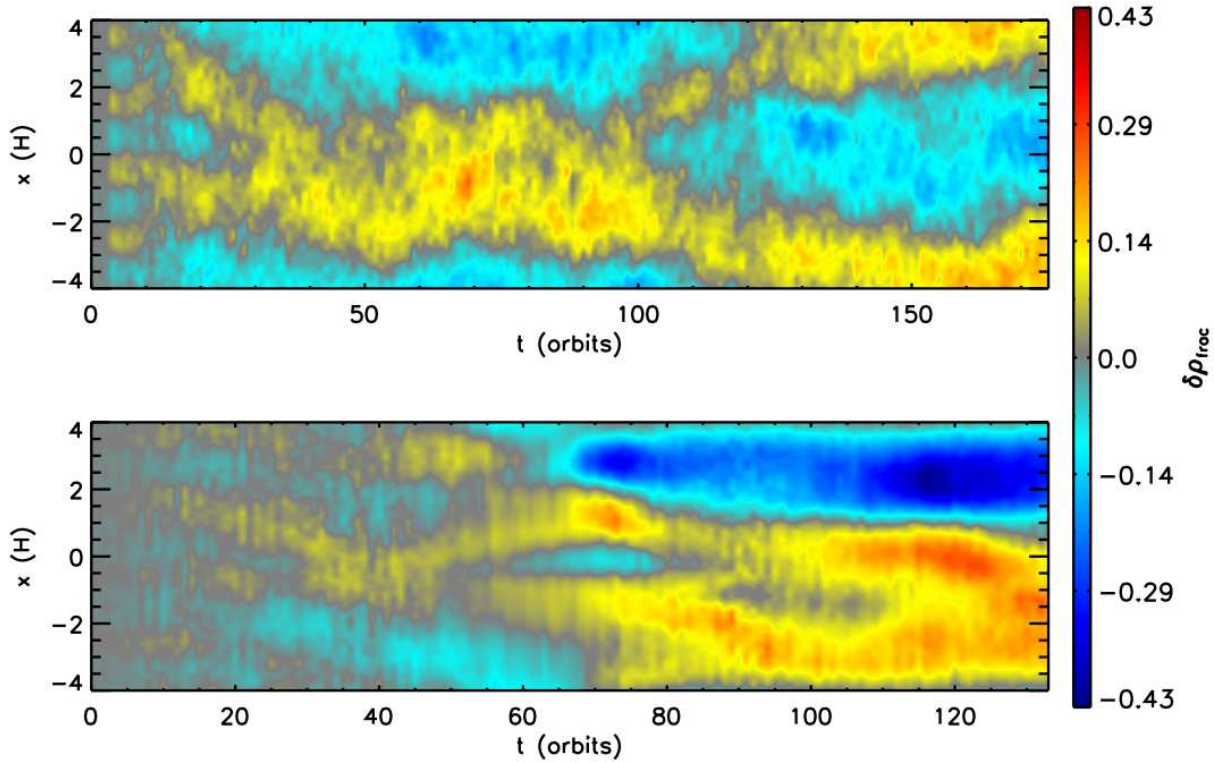}
\end{center}
\vspace{-0.1in}
\caption{
Space-time diagram of fractional gas density fluctuations, $\delta\rho_{\rm frac}$ (as defined in the text) in the $(t,x)$ plane for the ideal MHD run (IDEAL; top) and the ambipolar diffusion run (NF30AU; bottom).  For both runs, the vertical average is done for $|z| < 0.5H$, which is well within the ambipolar damping region for NF30AU.  The temporal axes on each plot are different.  The amplitudes of the zonal flows are comparable in the two simulations.
}
\label{sttx_d_multi1}
\end{figure*}
 
 \begin{figure*}[ht]
\begin{center}
\includegraphics[width=0.94\textwidth,angle=0]{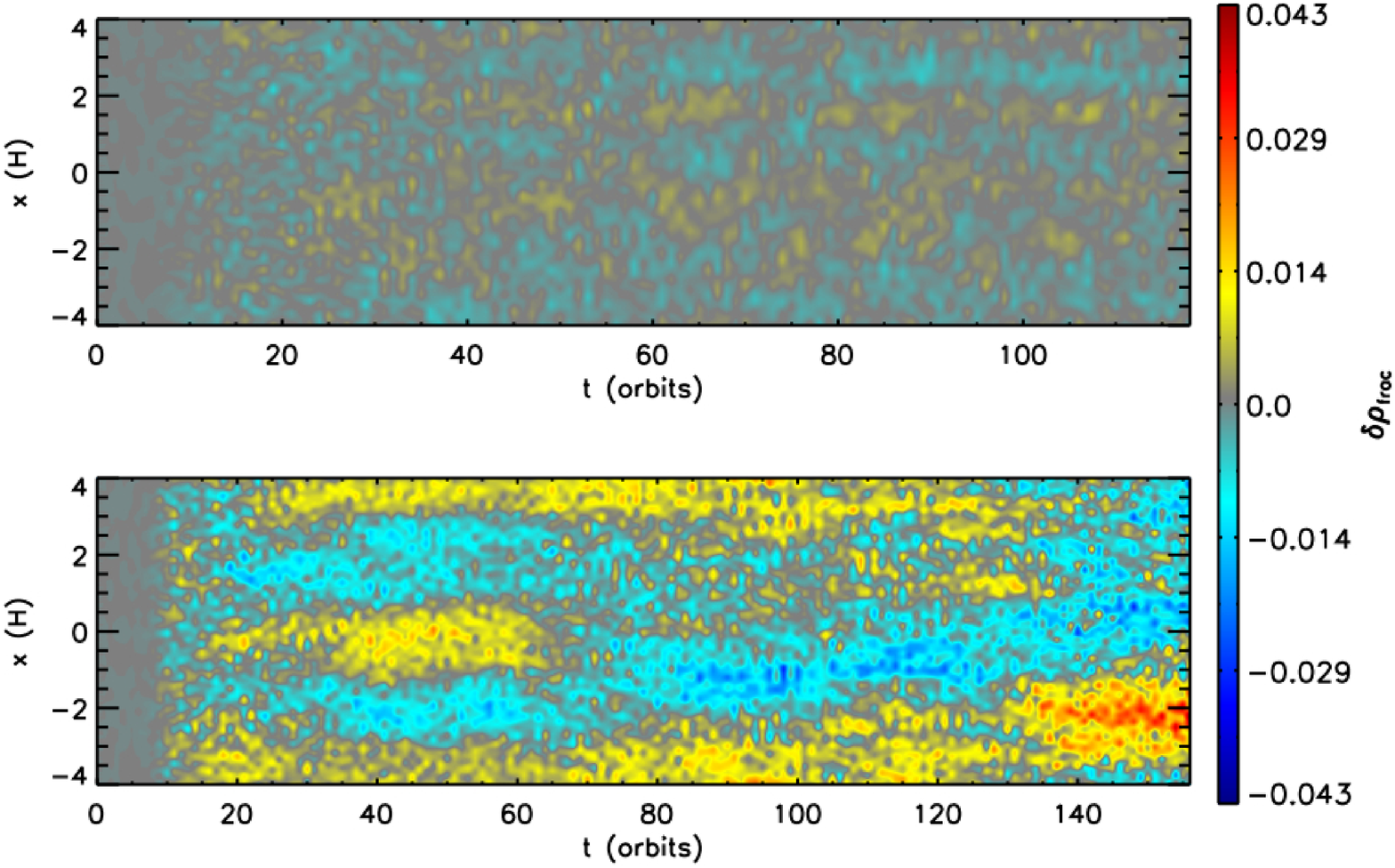}
\end{center}
\vspace{-0.1in}
\caption{
Space-time diagram of fractional gas density fluctuations, $\delta\rho_{\rm frac}$ (as defined in the text) in the $(t,x)$ plane for the zero net vertical magnetic flux runs with ambipolar diffusion at 30 AU (ZNF30AU; top) and 100 AU (ZNF100AU; bottom).  As with Fig.~\ref{sttx_d_multi1}, the vertical average is done for $|z| < 0.5H$.  The temporal axes on each plot are different.  For ZNF30AU, the zonal flows appear to be weak at best with a radial structure that is not well-defined.  The amplitude of these fluctuations is roughly an order of magnitude smaller than that of the zonal flows in ZNF100AU, which are themselves an order of magnitude weaker than both IDEAL and NF30AU (see Fig.~\ref{sttx_d_multi1}).  
}
\label{sttx_d_multi2}
\end{figure*}

The domain size for all of our simulations is $8H~\times~16H~\times~8H$ in $x,y,z$, 
where the vertical scale height $H$ is defined in terms of the sound speed $c_s$ and angular velocity $\Omega$ 
via $H = \sqrt{2} c_s / \Omega$. We parameterize the net vertical flux threading the simulation volume in 
terms of a parameter $\beta_0$, defined as the initial ratio of gas to magnetic pressure at $z=0$. 
All of our simulations use a resolution of 36 zones per $H$ and employ a density floor that is 
set to $10^{-4}$ in units of the initial mid-plane density to prevent prohibitively small time steps.
 
Our fiducial simulation is a variant of the AD30AU1e4 run of \cite{simon13b}. In this run, the ionization profile
is calculated using the above prescription at a radial location of 30~AU in the MMSN model.
The domain is threaded with a net vertical field characterized by a mid-plane gas to magnetic pressure ratio of $\beta_0 = 10^4$.
As in \cite{simon13b}, we impose upon this net field a sinusoidally varying vertical field in order to suppress the strong channel modes that develop
and disrupt vertically stratified simulations in the presence of a uniform vertical field \cite[see][]{miller00}.  However, unlike \cite{simon13b}, we initialize
this sinusoidal field at higher radial frequency so that the net vertical field is

\begin{equation}
\label{initi_b}
B_z = B_0 \left[1+\frac{1}{4}{\rm sin}\left(\frac{8\pi}{L_x}x\right)\right],
\end{equation}

\noindent
where $L_x$ is the domain size in the $x$ dimension, and $B_0 = \sqrt{2P_0/\beta_0}$ ($P_0$ is the initial, mid-plane gas pressure).  This is
done in order to distinguish the radial wavelength of the zonal flows that are produced (which is $\sim L_x$) from possible persistent artifacts 
that might result from the initial sinusoidal component of $B_z$. 
We have compared the volume-averaged stress values between this run and the
equivalent run in \cite{simon13b} and find excellent agreement. We label this run NF30AU.

We include here three additional runs for comparison.   
To explore the effect of a net vertical field, we have run a model with zero net vertical field but with all other
parameters the same as in NF30AU.  This run is labelled ZNF30AU (ZNF meaning ``zero net flux").   We have also run a
zero net vertical field model at 100 AU in the MMSN disk; this run is labelled ZNF100AU.  In both cases, the initial
magnetic field has a net toroidal geometry and decreases in strength away from the mid-plane so that $\beta = 100$
throughout the domain.\footnote{This field geometry has been used in our previous zero net vertical flux simulations \citep{simon13a}, and we employ this same geometry here to be consistent with these previous setups.  While we believe it is unlikely that a field geometry that precludes a net vertical magnetic flux but does not necessarily have a net toroidal component will produce qualitative differences, this has yet to be shown definitively.}  Finally, to compare our fiducial simulation with one that has no ambipolar diffusion, we have
run an ideal MHD shearing box with an identical magnetic field structure and strength to the fiducial run; this
calculation is labelled IDEAL. All simulations are listed in Table~\ref{tbl:sims}.

\section{Zonal Flow Strength}
\label{results}

A primary goal of this work is to analyze these various shearing box simulations and examine the characteristics of
zonal flows, if they are present. Figure~\ref{sttx_d_multi1} shows the radial space-time diagram of the fractional
variation in the $y,z$-averaged gas density, $\delta\rho_{\rm frac}$, for IDEAL and NF30AU.   The $z$ average was done for $|z| < 0.5H$, well
within the ambipolar damping zone of all of the simulations that include ambipolar diffusion (we checked other
vertical domains over which to average; no significant differences were seen).  The figure shows the development of
zonal flows over long timescales and that these zonal flows have similar amplitudes in the damped region of NF30AU
compared to IDEAL.  Thus, despite there being reduced MRI stresses (by at least an order of magnitude) in the
ambipolar damping zone, zonal flows persist as strongly as they do in the fully ideal MHD case.

Figure~\ref{sttx_d_multi2} shows the same space-time diagnostic, but for the two simulations with no vertical magnetic field.  The amplitude on the color bar has been decreased by a factor of 10.  The strength of density fluctuations for ZNF100AU is roughly a factor of 10 lower than for IDEAL and NF30AU.  The radial scale of zonal flows in ZNF100AU evolves over time, but appears to ultimately end in a configuration that has the same scale as IDEAL and NF30AU.  The simulation ZNF30AU has even weaker zonal flows (by roughly another order of magnitude). The radial length of these flows is less well-defined but appears to be smaller than that of the other simulations.

\begin{figure}
\begin{center}
\includegraphics[width=0.5\textwidth,angle=0]{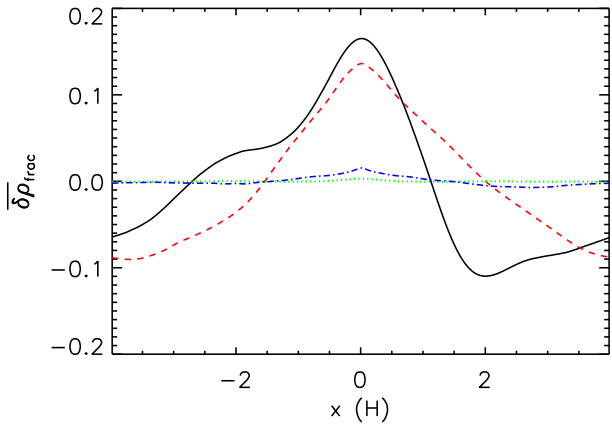}
\end{center}
\vspace{-0.1in}
\caption{Time-average of the fractional gas density fluctuation. Before time averaging, we first shifted the maximum of $\delta \rho_{\rm frac}$ to be at $x = 0$, as explained in the text.  The black solid line corresponds to NF30AU, red, dashed line to IDEAL, green, dotted line to ZNF30AU, and blue dot-dashed line to ZNF100AU.  The time average was calculated from orbit 50 to the end of each run.  The net vertical field run with ambipolar diffusion (NF30AU) and the ideal MHD run (IDEAL) both show strong density fluctuations of order 0.1--0.2.  The remaining runs contain a MRI-dead region, and here the fluctuations are comparatively very weak.
}
\label{sttx_d_avg}
\end{figure}

We further compare the amplitude of these zonal flows by time-averaging the fractional density fluctuation as shown in Fig.~\ref{sttx_d_avg}.  In this analysis, we shift the maximum of $\delta \rho_{\rm frac}$ to be at $x = 0$ at each time and then time-average the resulting profile from orbit 50 onwards. The figure emphasizes the previous point; the zonal flow amplitudes in IDEAL and NF30AU are quite similar, reaching $\sim 0.1$--0.2.. Without a net vertical field, the amplitude drops drastically. 

As found previously \citep{simon13a}, the ambipolar damping regions of these zero net vertical field simulations are
nearly completely devoid of turbulent activity.  In the absence of a vertical net magnetic field, the MRI is
completely quenched in this region, and any stress that is present results from Reynolds stresses induced by the
active layers as well as large scale correlations in the toroidal and radial magnetic fields.  Thus, in the absence
of a net vertical field, the ambipolar damping region is very similar to the Ohmic dead zone present in the inner
regions of disks \citep{gammie96,simon13b}.  Figures~\ref{sttx_d_multi2}~and~\ref{sttx_d_avg} clearly show that the zonal flow amplitudes in
these ``dead" regions are very small, reaching a maximum of $\sim 0.04$.

To determine whether or not there is a correlation between the stresses and the zonal flow amplitudes, we calculate the $\alpha$ value, defined here as

\begin{equation}
\label{alpha}
\alpha \equiv \overline{\left[\frac{\left\langle \rho v_x \delta v_y
- B_xB_y\right\rangle}{\left\langle \rho c_s^2\right\rangle}\right]},
\end{equation}

\noindent
where the angled brackets denote a volume average over the entire domain, and the bar over the ratio denotes a time-average from 50 orbits onward.   We also perform the same calculation, but within the region $|z| < 0.5H$; this quantity is defined as $\alpha_{\rm midplane}$. All of these calculated quantities are displayed in Table~\ref{tbl:sims}.  The quantity $A$ is the maximum absolute value of the time-averaged fractional density fluctuation from Fig.~\ref{sttx_d_avg}.  

The runs IDEAL and NF30AU have very similar values for $A$ ($\sim$0.1--0.2), though they have significantly different stress values (both $\alpha$ and $\alpha_{\rm midplane}$).  There may be a correlation between the values of $\alpha_{\rm midplane}$ and $A$ for ZNF30AU and ZNF100AU.  However, from only these two data points, one cannot draw any firm conclusions.  We also compared the time-averaged strength of the toroidal field within the mid-plane region to the zonal flow amplitude, finding similarly inconclusive numbers.  If there is any correlation between the strength of the stress or magnetic field and the zonal flow amplitude, such a correlation is not present from this work.  A further exploration of possible relationships between turbulence levels and zonal flow amplitudes would require significantly more simulations; this is beyond the scope of our current work but will be addressed in future publications.  

The results of \cite{simon13a,simon13b} suggest that in order for MRI turbulence to induce accretion rates that agree with observational constraints, the outer disk regions must be threaded with a relatively weak ($\beta_0 \sim 10^4$--$10^5$) vertical magnetic field.  Our results here show that in this limit, zonal flows are indeed present in the outer regions of protoplanetary disks, even in the region where the MRI is most strongly damped due to ambipolar diffusion. 

These strong zonal flows are present near the disk mid-plane where planet formation is likely to take place.  However, are these zonal flows sufficiently strong to trap small particles, potentially inducing planet formation processes?  We address this issue in the next section.

\section{Particle Trapping}
In our local simulations the mean surface density is uniform in the $x$ (radial) direction. Any positive density perturbation 
then suffices to create a local pressure maximum. In a global 
disk model the situation is less clear cut. Commonly considered models have steeply declining profiles of mid-plane pressure, so 
that depending upon their radial scale quite substantial perturbations may be required before any pressure maximum is present. 
It is not entirely clear how to translate local estimates of zonal flow amplitudes into global predictions for particle trapping. 
Here we adopt the simplest approach. We evaluate $(1 + \delta \rho_{\rm frac}) (r)$ from the simulations, and assume that the 
corresponding global profile is the mean mid-plane pressure multiplied by this function.

Within this framework, we use two methods to determine whether particles would be trapped within our zonal flows. 
First, we model these flows as sinusoidal functions with amplitude $\epsilon$.
Such a model is a reasonable approximation given that the shape of these zonal flows resembles a sinusoidal function (see
Fig.~\ref{sttx_d_avg}). Furthermore, this approximation has previously been used to represent zonal flows \citep{pinilla12}.  We consider the
radial pressure profile of a model disk with surface density proportional to $r^{-q}$ ($q = 1.5$ for the MMSN) and add a sinusoidal
perturbation to give us

\begin{equation}
\label{pressure}
P(r) = C r^{-q-7/4}\left(1+\epsilon {\rm cos}\left[\frac{2\pi(r-r_o)}{L(r)}\right]\right)
\end{equation}

\noindent
where $C$ is an arbitrary constant, $r_o$ is the center of the zonal flow (i.e., $r_o$= 30AU for NF30AU), and

\begin{equation}
\label{length}
L(r) = \delta H = \delta (0.042) r^{5/4}
\end{equation}

\noindent 
is the width of the zonal flow, expressed as the number of scale heights $H$ via the dimensionless number
$\delta$.  The far right-hand-side of equation~(\ref{length}) results from assuming a disk thickness 
$H / r = 4.2 \times 10^{-2} (r/ {\rm AU})^{1/4}$, roughly consistent with a MMSN model\footnote{Note that our definition of
scale height $H$ is a factor $\sqrt{2}$ larger than that defined by $h = c_s / \Omega$.}.

\begin{figure}
\begin{center}
\includegraphics[width=0.5\textwidth,angle=0]{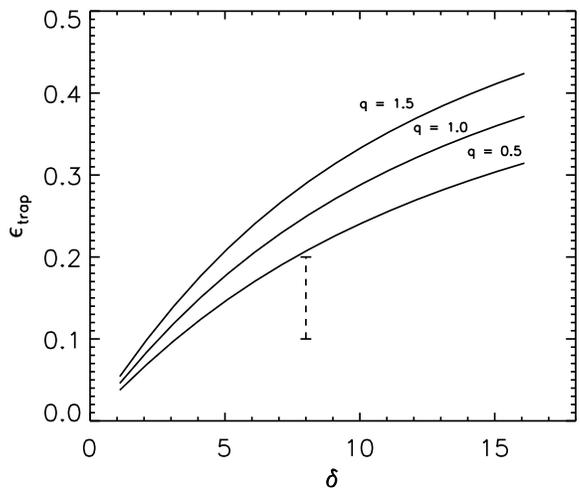}
\end{center}
\vspace{-0.1in}
\caption{The minimum amplitude, $\epsilon_{\rm trap}$, of a sinusoidal zonal flow such that $dP/dr = 0$ near (but
not at) $r_o$ plotted as a function of the width of the zonal flow, $\delta$, in units of $H$.  For values of $\epsilon >
\epsilon_{\rm trap}$, $dP/dr > 0$ near $r_o$ and strong particle trapping is possible.  The different curves are labelled by
their surface density slopes, with $q = 1.5$ corresponding to the MMSN model.  The vertical dashed line spans the
approximate range in zonal flow extrema inferred from our primary simulation NF30AU ($A \sim$ 0.1--0.2; see
Fig.~\ref{sttx_d_avg}). For typically assumed mean surface density profiles, the amplitude and radial scale 
of the simulated zonal flows marginally fail to satisfy the condition for particle trapping, even given a 
net vertical magnetic field.}
\label{eps_delta}
\end{figure}

We solve equation~(\ref{pressure}) to determine the minimum value of $\epsilon$, named $\epsilon_{\rm trap}$, such that $dP/dr = 0$ near $r_o$;  for $\epsilon > \epsilon_{\rm trap}$, particle trapping is possible.  The result is shown in Fig.~\ref{eps_delta} for several surface density profiles (i.e., different $q$ values).  The minimum $\epsilon$ necessary to trap particles is plotted against the typical length scale (in units of $H$) of a zonal flow.  As the width of the zonal flow increases, a larger amplitude is required in order to trap particles; this result is consistent with the work of \cite{pinilla12}. 

The vertical dashed line corresponds to the width of the zonal flow in our primary simulation NF30AU, which is $8H$, and a range in amplitudes for the time-averaged zonal flow as shown in Fig.~\ref{sttx_d_avg}.  These results suggest that particle trapping is only possible for $\epsilon_{\rm trap} \gtrsim 0.2$ at $\delta = 8$. The time-averaged amplitude of the zonal flow in NF30AU is smaller than this trapping value, suggesting that to first order, particle trapping is {\it not} likely even with the creation of strong zonal flows in the outer regions of disks threaded with a vertical magnetic field. 

Recent studies \cite[e.g.,][]{dzyurkevich10,dittrich13,simon12} suggest that the width of zonal flows is roughly somewhere between $5H$ and $10H$ (with the exception of \cite{uribe11}, whose results suggest a slightly larger zonal flow width).  Since $\epsilon_{\rm trap}$ is a shallow function of $\delta$, our conclusions do not change significantly when considering this larger range of possible widths.

We also examine the radial pressure gradient with the perturbation of NF30AU added directly to the pressure function, rather than assuming it to be a sinusoidal perturbation.  We start with equation~(\ref{pressure}), and add the function $\overline{\delta\rho}_{\rm frac}(x)$ in place of the cosine term.  

\begin{equation}
\label{pressure2}
P(r) = C r^{-q-7/4}\left[1+\overline{\delta\rho}_{\rm frac}(x)\right]
\end{equation}

\begin{figure}
\begin{center}
\includegraphics[width=0.5\textwidth,angle=0]{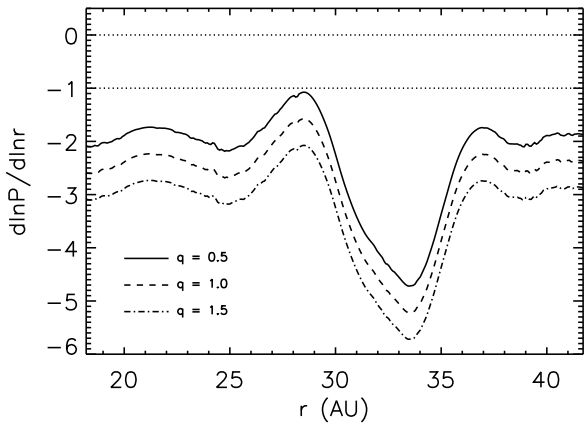}
\end{center}
\vspace{-0.1in}
\caption{Logarithmic radial pressure gradient calculated by imposing the time-averaged zonal flow structure from run NF30AU onto several model pressure profiles (labelled via the $q$ parameter).  The two horizontal, dotted lines from top to bottom correspond to ${\rm d ln}P/{\rm dln}r = 0$ and ${\rm d ln}P/{\rm dln}r =-1$. The peak in the gradient is located at $r \sim 28$AU, but never reaches ${\rm d ln}P/{\rm dln}r = 0$, suggesting that particles will not be trapped by the zonal flow present in our simulation.
}
\label{dpdr}
\end{figure}

\noindent
Our radial shearing box coordinate $x = r-r_o$.  We choose $C = 1$ without loss of generality, and for this run, $r_o = $30AU.  We plot the logarithmic radial gradient of the pressure in Fig.~\ref{dpdr}, assuming three different values for $q$ as in Fig.~\ref{eps_delta}.  The logarithmic radial pressure gradient reaches a maximum of ${\rm d ln}P/{\rm dln}r \sim -1$ in the most optimistic scenario ($q = 0.5$).

 Our results suggest that given current best estimates of the width and amplitude of ambipolar zonal flows, 
the resulting perturbations are likely too weak to strongly trap particles in the outer regions
of protoplanetary disks. This remains true even with the enhancing effect of a net vertical magnetic field, 
which substantially increases the vertically integrated angular momentum transport rate (bringing it up to levels 
consistent with measured accretion rates). We note, however, that the inferred failure to trap particles involves 
a mismatch between what is measured and what is needed by a factor that is only $\sim 2$. A perturbation 
whose amplitude is twice as large, or whose radial scale is a factor of two smaller, would potentially 
trap particles.

\vspace{0.3in}
\section{Discussion}

Our primary result is that while we see strong zonal flows, even in the presence of damped turbulence from ambipolar diffusion, the
amplitudes obtained from our simulations are not sufficiently large to create pressure maxima where particle trapping can occur.  That the
strength of zonal flows in the presence of ambipolar diffusion is similar to that in the ideal MHD limit is surprising.  
Further examination of the relationship between turbulence levels and zonal flow amplitudes will require a larger
parameter study than that performed here.

Our conclusions are less optimistic for the prospects of particle concentration than 
those of \cite{pinilla12} who find that particle trapping {\it can} occur in their model disks. 
The reason for this difference is the assumed width of the zonal flows, which \cite{pinilla12} 
take to be comparatively narrow ($\delta \sim 1$). From Fig.~\ref{eps_delta} it is 
clear that such a narrow width, if combined with our measured amplitudes, would indeed lead to 
particle trapping for any reasonable background surface density profile. The problem with this scenario 
is that our simulations, along with several previous works \citep{johansen09,simon12}, suggest 
that $\delta \sim 5-10$ is more realistic. With these parameters, trapping particles directly 
by creating local pressure maxima is significantly more difficult. 
 
There remain several uncertainties in our work.  First, while our simulations include the key MHD physics present in the outer regions of
protoplanetary disks (i.e, strong ambipolar diffusion), we have only run a limited number of these simulations,  for a relatively short
interval. The roughly 100 orbit duration of our runs is limited by their computational demands,  which remain prohibitive despite the
considerable speed up from the use of super-time-stepping.   The bottom panel of Fig.~\ref{sttx_d_multi1} shows that towards the end of the
calculation, the fractional variation in gas density approaches $\sim 0.3$, and it is possible that integrating this run further would produce
sustained zonal flows of roughly this amplitude.  However, it is also possible that these particular zonal flows will decrease in strength
and then fluctuate in a stochastic manner, as is observed in IDEAL and previous shearing box simulations \cite[e.g.,][]{johansen09,simon12}.
Second, there are some uncertainties in translating the local simulation results to predictions for trapping in global
disk models. High resolution global disk simulations that include ambipolar diffusion, although currently challenging to run, will be
essential for a definitive determination of whether outer disk zonal flows can trap particles. 
Finally, while at face value our results show that particles will not be {\it trapped}, the pile-up effect due to changes in the radial drift velocity
may still play a role in the planet formation process \citep{johansen06,dittrich13}.  We plan to address these various uncertainties in future
work. 

Even if zonal flows fail to strongly concentrate particles, their presence may still be observable. 
At 30~AU, $H / r \approx 0.1$, and the zonal flow widths found in our simulations correspond to 
physical scales of the order of 5~AU. Although the perturbations to the gaseous surface density 
are modest, some degree of enhancement of the particle density due to differences in the radial drift 
speed across the flow is expected \citep{johansen06,dittrich13}. Early ALMA results show that there are 
some disks whose outer regions display dramatic {\em non-axisymmetric} dust distributions \citep{vdm13}, 
and the relatively subtle axisymmetric structures resulting from our zonal flows would likely not be 
detectable in such systems. It may, however, be possible to observationally constrain the existence and 
properties of zonal flows by focusing on those disks with smooth radial profiles and the 
smallest departures from axisymmetry.

\acknowledgments

We thank our collaborator Xuening Bai for his work on the ambipolar disk simulations and Rebecca Martin and Jeff Oishi for useful
comments regarding this work.  We also thank the anonymous referee, whose suggestions greatly enhanced the quality of
this work. We acknowledge support from NASA through grant NNX13AI58G, from the NSF through grant 
AST 1313021, and from grant HST-AR-12814
awarded by the Space Telescope Science Institute, which is operated by the  Association of Universities for Research
in Astronomy, Inc., for NASA, under contact NAS 5-26555. J.B.S.'s support was provided in part under contract with
the California Institute of Technology (Caltech) funded by NASA through the Sagan Fellowship Program executed by the
NASA Exoplanet Science Institute. This research was supported by an allocation of advanced computing resources
provided by the National Science Foundation. The computations were performed on Kraken and Nautilus at the National
Institute for Computational Sciences through XSEDE grant TG-AST120062.

\end{document}